# Synthetic dataset generation methodology for Recommender Systems using statistical sampling methods, a Multinomial Logit model, and a Fuzzy Inference System.

Vitor T. Camacho[1]

[1] PhD, vitor.camacho@syone.com, R&D Data Science, Syone.

*Abstract*

It is said that we live in the age of data, and that data is ubiquitous and readily available if one has the tools to harness it. That may well be true, but so is the opposite. It is ever more common to try to start a data science project only to find oneself without quality data. Be it due to just not having collected the needed features, or due to insufficient data, or even legality issues, the list goes on. When this happens, either the project is prematurely abandoned, or similar datasets are searched for and used. However, finding a dataset that answers your needs in terms of features, type of ratings, etc., may not be an easy task, this is particularly the case for recommender systems. In this work, a methodology for the generation of synthetic datasets for recommender systems is presented, thus allowing to overcome the obstacle of not having quality data in sufficient amount readily available. With this methodology, one can generate a synthetic dataset for recommendation composed by numerical/ordinal and nominal features. The dataset is built with Gaussian copulas, Dirichlet and Gaussian distributions, a Multinomial Logit model and a Fuzzy Logic Inference System that generates the ratings according to different user behavioural profiles and perceived item quality.

## 1    Introduction

Nowadays, data is ubiquitous, and it is frequently said that we live in the age of data, so much so that concepts like big data, machine learning, AI, have become almost void of meaning due to the frequency with which they are uttered. However, even though data is everywhere, quality data or the data that we need for a given end, or more importantly, data we can use, is not always available. It is ever more common to struggle in the beginning of data science/machine learning projects with the non-availability of the necessary data. This is common and happens for several reasons. Some of the most common reasons, for example, some needed features might not have been collected beforehand; or the available data cannot be used due to legal reasons; insufficient or non-existing data altogether, etc.

When the necessary data is not present, one can choose from a series of alternatives: either the project is abandoned; or resources are spent collecting the necessary data, which isn't immediate and may not be feasible; or alternative datasets are used, which will likely be somewhat different in terms of features and statistical properties than the would-be data for the project. This is particularly difficult for recommender systems where you need many different datasets corresponding to user data, user-item interaction, item data, etc. A last solution is generating synthetic data. The advantages of generating one's synthetic data is that one can choose the statistical properties of the data (distributions, correlation matrix, etc), the sparsity of ratings, the type of ratings (unary, binary or 5-stars), etc. The upside is more control over the data and the ability to test recommendation algorithm's (ex.: factorization machines) ability in identifying latent features that one has purposely introduced in the dataset, for example. The downside is that it is not an easy task to generate good quality datasets for recommender systems and it may also be the case that you don't exactly know what statistical properties, correlations, latent features you should encode into the synthetic data.

In this work, a methodology for the generation of synthetic datasets for recommender systems is presented, thus allowing to overcome the obstacle of not having quality data in sufficient amount (or even at all) readily available. The difficulties that are associated with this task are essentially the definition of a dataset with multiple datatypes, such as numerical (continuous), ordinal and nominal, and with different levels of correlation among the data, as well as the definition of user-ratings based on well-defined latent user preferences. For this, a methodology was devised where several different techniques are employed in sequence to create the datasets concerning user characteristics, item properties, item categories and latent user preferences associated to user and item features, and as a result, a user-item sparse ratings matrix. The output of the methodology is:

1) Item dataset with item names and categories.

2) User dataset with user characteristics (demographic features).

3) User-item sparse ratings matrix.

4) Latent preferences and Multinomial Logit model to compare with the outputs of the Recommender System.

## 1.1 Literature Review – Related Work

There are several works in the available literature that present different approaches, usually not for the explicit formulation of a dataset for machine learning or recommender system training. Additionally, many of the works don't present approaches with multiple datatypes, particularly for the generation of dependent nominal and ordinal features. However, the main difference for the rest of the literature is probably the definition of a user preference vector based on a Multinomial Logit defined with the previously composed user characteristic dataset and employing a Fuzzy Inference System to define the user-item ratings.

Relative to Recommender Systems, the literature is full of examples of application of Recommender Systems to given industries, particularly the tourism industry, which relates to the dataset case study in this work [1]–[7]. In these works, two refer to an overview of techniques employed in tourism recommender systems [7] and a survey of the existing literature on that theme [1]. The other referenced works contain some examples of recommender systems designed for tourism, with particular emphasis given to the Sem-Fit [4] and SigTur [3] projects.

As already mentioned, regarding the generation of synthetic datasets, there are a few works that have developed on this topic, which are presented next. It is important to mention that this doesn't intend to be a comprehensive collection of works on this theme [8]–[21]. It does nonetheless contain relevant and relatively recent works, for the most part, on the theme of synthetic data generation using different methods and for various applications. In this collection, there are a series of works for the generation of datasets for recommender systems, such as the example of Suglia et al. [10] with a procedure for the generation of datasets for conversational recommender systems. In Rodriguez-Hernandez et al. [8] a synthetic dataset generator DataGenCARS is presented. This generator simulates datasets for context-aware recommender systems and has many of the same properties as the one presented in this work, but the methods employed are not as transparently explained as in this work. In Pasinato et al. [9], another method for the generation of synthetic data for context-aware recommender systems is presented. In this case they generate data based on users' and destinations' profiles which are modeled by random variables and their PDFs. In Monti et al. [12], a clustering approach is applied for the generation of datasets for recommendation. In this approach synthetic datasets are created from a configurable number of users that mimic the characteristics of already existing users. In Belletti et al. [15] a method that adapts Kronecker Graph Theory is presented which produces recommendation datasets through fractal expansions. Other works are also presented besides ones whose goal is the generation of datasets for recommender systems. For example, two works present R packages for the generation of correlated variables. In Amatya et al. [13], an R package is presented designed for the concurrent generation of correlated ordinal and normal data. In Tsoukalas et al. [14] an R package anySim is presented, which can be used to generate non-Gaussian correlated random variables, stochastic processes and random fields. Among these works there are some that present methods for the simulation of correlated variables using multivariate distributions and multivariate categorical variates [17], [18]. Another work by Triastcyn et al. [16] presents a method for the generation of artificial data for private deep learning, and in Surendra et al. [11] a review of synthetic data generation methods for privacy preserving data publishing is given. Finally, the works of Asre et al. [21], Astolfi [20], and Xu et al. [19] present methods for the generation of tabular data using generative adversarial networks (GANs).

One of the novel characteristics of this work compared to the literature is in the methods employed for dealing with dependent ordinal/numerical features and nominal features, particularly regarding correlation. Another novel characteristic is employing a Multinomial Logit model to generate latent features associated to user preference. Finally, a Fuzzy Inference System is employed where the latent features, the item categories, and user behavioural traits and implicit item quality are used to generate ratings in a fuzzy logic framework. In the literature there are many works that may achieve similar outputs as this one, however some are not very clear in the methods they employ and most use different methods than the ones presented herein, particularly the use of Multinomial Logit for latent user preferences and the use of those preferences, user behaviour traits and implicit item quality as inputs of the Fuzzy Inference System with an output/consequent of the system being the user ratings.

## 1.2 *Techniques Used for Feature and Ratings Generation*

As mentioned before, in this work different techniques are used for the numerical/ordinal and nominal data. For the generation of the numerical/ordinal data, Gaussian copulas are used, with the definition of a covariance matrix that models the covariance between the numerical/ordinal features. In the case of the nominal features, these are generated by sampling from a Multinomial distribution with Dirichlet priors. After having defined the user features, the user preferences are generated. To generate the user latent preferences a Multinomial Logit model is used, and a user-preference matrix is derived. Finally, to generate the user ratings of the items, a Fuzzy Inference System is developed.

In the following sub-sections, a theoretical overview is given for each one of the employed techniques.

### 1.2.1 *Gaussian Copulas*

As mentioned previously, Gaussian copulas are employed for the generation of the ordinal/numerical attributes. Gaussian copulas are just one type among many copulas, such as the *t* copula, bivariate Gumbel copula, bivariate Clayton, etc., which are used particularly in quantitative finance to model tail risk and for portfolio-optimization applications [22][23], but also in fields such as structural engineering, reliability engineering, and any field where simulation of dependent variables is to be done [11], [24], for example, through Monte Carlo Simulation.

The definition of a copula is a multivariate distribution function for which the marginal probability distribution of each variable is uniform on the interval [0,1]. Copulas are used to model dependence between random variables. The Gaussian copula can be defined mathematically as follows. Let $X \sim MN_d(0, P)$, where P is the covariance matrix of X. The corresponding Gaussian copula is defined as

$$C_P^{Gauss}(u) \coloneqq \Phi_P(\Phi^{-1}(u_1), \dots, \Phi^{-1}(u_d))$$

Where $\Phi(\cdot)$ is the standard univariate normal CDF and $\Phi_P(\cdot)$ denotes the joint CDF of $X$. The steps for its simulation are given in 3.2.1.

## 1.2.2 *Multinomial Distribution with Dirichlet Priors*

To simulate nominal random variables, different methodologies should be used because nominal features don't have a hierarchical ordering, as do ordinal and numerical features. This means that one cannot define any of the nominal feature's labels as being higher or lower in value than another label. Therefore, sampling the same way as with ordinal features, via Gaussian copulas, would encode monotonical feature correlation, which would imply ordering. Thus, to simulate these features a different approach was devised where a Multinomial distribution is employed, together with Dirichlet priors, that define each features Multinomial distribution parameters according to a given sample of the population. These samples of the population are defined as α values.

The process can be described by the following expressions:

$$p(\theta|\alpha) \propto \prod_{j=1}^{k} \theta_j^{\alpha_j - 1}$$

$$\theta_j \geq 0 \text{ and } \sum_{j=1}^{k} \theta_j = 1$$

Where

$$\begin{cases} \alpha \sim some\ distribution\ (sample\ values) \\ \theta \sim Dirichlet_K(\alpha) \\ Z \sim Multinomial_K(\theta) \end{cases}$$

These expressions are applied in section 3.2.3 to generate synthetic nominal features for every user, given α values as input.

## 1.2.3 *Multinomial Logit*

The link between the users and the items is given by user preferences, which can be modelled as category preferences which then map to item preferences given the items categories. One way to define these preferences, and the resulting user preference vector is via a Multinomial Logit model. The way a Multinomial Logit model works is through the definition of utility functions that map user features to preferences via β values, which describe how each user feature or item feature affects the utility score of an item to a user. There are a few works that present R packages for the estimation of multinomial logit models [25], [26]. The mathematical description of how a Multinomial Logit model works is given in the following.

$$u_{ij} = v_{ij} + \varepsilon_{ij} \sim Extreme\ Value\ Type\ I\ (Gumbel)$$

$$y_i = argmax_{j \in \{1,...,J\}} u_{ij}$$

$$v_{ij} = x_i \beta_j$$

Where *i* – individual user; *j* – (preferences) alternatives/categories; *k* – characteristics of the user; *ε* – error terms representing utility that is not explained by the features/model; *v* – utility explained by the model; *u* – utility of alternative *j* to user *i*.

Characteristics of individual user i: $x_{ik}$

Utility dependence on $x_k$ for alternative j, relative to alternative 1: $\beta_{jk}$

A particular feature of the Multinomial Logit model is that it is easy to attain the conditional probabilities of user *i* preferring category/alternative *d*, compared to all alternatives/categories *J*, through the following expression:

$$\Pr(d|v_{ij}) = \frac{\exp(v_{id})}{\sum_{j=1}^{J} \exp(v_{ij})}$$

It is through this expression that the utility values are mapped to conditional probability values. In sections 3.2.4 and 3.2.5, it is shown how the Multinomial Logit is used to simulate data and how the conditional probabilities are calculated, which are then used as input to the Fuzzy Inference System.

### *1.2.4 Fuzzy Inference System*

A Fuzzy Inference System (FIS) is a system that uses fuzzy set theory [27] to map inputs to outputs. It is the key unit of a fuzzy logic system having decision making its primary work. It uses "IF… THEN" rules along with connectors "OR" or "AND" for drawing essential decision rules. A FIS can be defined as essentially having five functional block units:

- Rule Base – Contains fuzzy IF-THEN rules
- Database – Defines the membership functions of fuzzy sets used in fuzzy rules.
- Decision-making Unit – Reasoning mechanism that performs the induction made upon the guidelines and the inputs given to infer the output.
- Fuzzification Inference Unit – Converts the crisp quantities into fuzzy quantities.
- Defuzzification Inference Unit – Converts fuzzy quantities into crisp quantities.

The following figure shows a flowchart that represents how these components interact with each other and how a FIS works.

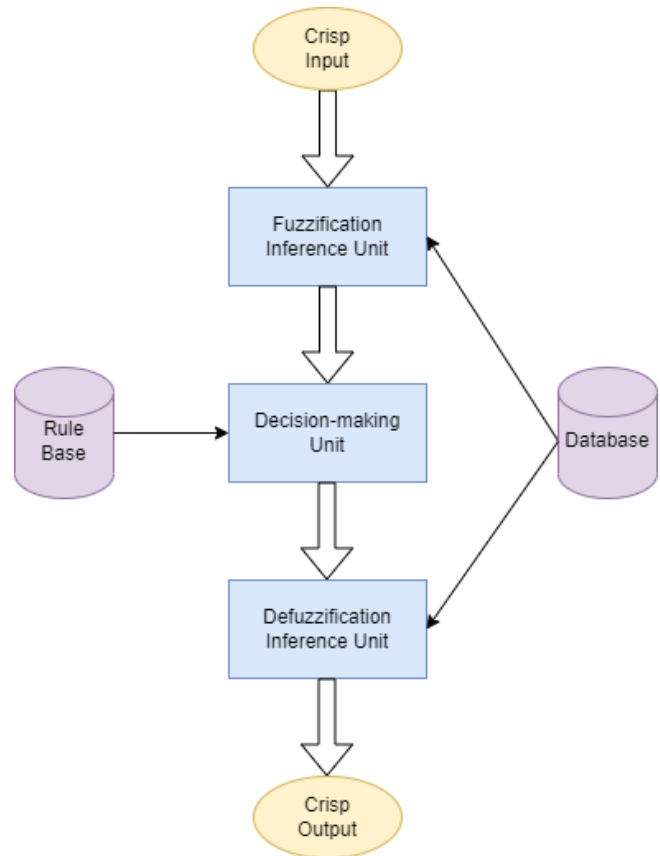

*Figure 1 Flowchart showing how a generic FIS works.*

There are essentially two important methods/variants of FIS, the Mamdani Fuzzy Inference System [28] and the Takagi-Sugeno Fuzzy Model (TS Method) [29]. In this work, the former is applied. The main difference between both methods is essentially related to how the output is given. In the case of the Mamdani FIS approach there is an output membership function, and the crisp result is obtained through defuzzification of the rules' consequent. In the case of the Takagi-Sugeno FIS approach, there is no output membership function and thus no defuzzification step, but rather a weighted average of the rules' consequent which are already crisp results since the Takagi-Sugeno rules are of the shape, IF x is A AND y is B THEN z=f(x,y). Thus, the Takagi-Sugeno approach maps the inputs directly to a crisp result through a mathematical function, while the Mamdani approach, on the other hand, maps the inputs to a fuzzy logic set, which then needs defuzzification to return a crisp result.

As mentioned before, the Mamdani FIS approach is applied herein, and more details on its application are provided in sections 3.2.7, where the application of step 10 of the methodology is done, with the presentation of the different fuzzy sets and a sample of the fuzzy rule base.

*1.3 Structure of the paper*

In this first section, the introduction was given plus an overview of the techniques used and relevant literature. In the second section the methodology is presented step-by-step, in a total of 10 steps, with a brief description and the output of the methodology. In the third section, the presented methodology is applied for the generation of a synthetic dataset to develop and train a recommender system for the tourism industry. In the beginning of that section the dataset and its features are described, particularly the different datatypes that compose it. Finally, in section 4, conclusions and future works are discussed.

**2 Data Generation Methodology**

In the following the methodology is presented with a step-by-step walkthrough of the entire data generation process. Each of the steps presented is later applied for the generation of a dataset for a tourism recommender system.

1. Sample continuous correlated latent processes (ordinal variables) - define covariance matrix.

2. Define alphas (cut-off) for ordinal variables.

3. Define alphas for each category in each nominal variable for the Dirichlet distributions.

4. Sample from the Dirichlet distribution to obtain theta values and sample from the Multinomial distribution with the sampled Dirichlet priors (theta) - sampling of the nominal variables.

5. Define Multinomial Logit model for preferences by defining beta values for all preferences according to all individual characteristics (both nominal and ordinal variables)

6. Calculate the probabilities of each user preferring each category, thus defining each user's preference vector. Define the user preference ($U_{Pref}$) matrix with all individual preference vectors.

7. Define a list of items with corresponding categories and build item category (icat) matrix.

8. Perform the external product of the $U_{Pref}$ matrix with the transposed icat matrix to obtain user-item matrix.

9. Add noise to the user-item matrix by adding a random sparse matrix to the user-item matrix with a given sparsity density. This adds additional noise in the form of individual item preferences that are not explained by the model (slightly different from the noise in the multinomial logit model which models category preferences that are not explained by the model).

10. A Fuzzy Inference System is employed where the users are defined according to two behavioural traits regarding how they rate items, and items are defined according to an implied value of overall quality (good, average, or bad). Then the Fuzzy Inference System defines the rating given by each user to each item, taking into account the user behavioural traits, the item implied quality and the user-item matrix, which has the likelihood of each user liking each item. The resulting user-item rating matrix is populated with the result of the Fuzzy Inference System according to a sparsity measure which defines the percentage of non-null entries of said matrix.

The output is: 1) item matrix with item names and categories; 2) user matrix with user characteristics; 3) user-item ratings matrix; and 4) latent preferences matrix and Multinomial Logit model to compare with the outputs of the recommender system algorithms, particularly latent factors of field-aware factorization machines. In the following section, this methodology is applied for the generation of a synthetic dataset.

## 3    Data Generation Results of the Case Study Application

In this section, the methodology presented in the previous section is applied to a case study of a Recommender System for the tourism industry. In the next sub-section, the various features and datatypes of the various data frames are presented. The data generation results are then presented in a step-by-step manner according to the steps of the methodology.

### *3.1    Feature definition of a tourism related dataset*

As said before, this work focuses on the generation of datasets for the training of a recommender system. In this case, specifically for a recommender system in the field of tourism, particularly a dataset on the services offered by a hotel to its clients.

Before presenting the methodology for the data generation, the features and feature types are defined. The features vary based on the data set they belong to and the type of features they are. Concerning the data set, the features can be divided in the following groups:

- Demographic Features
- Preferences
- Item Features
- User Ratings

Where,

- Demographic Features:
    - User ID
    - Age
    - Gender
    - Job
    - Academic Degree
    - Budget
    - Country/Region
    - Group Composition
    - Accommodation
- Preferences:
    - User ID
    - Beach
    - Relax
    - Shop
    - Night Life
    - Theme Park
    - Gastronomy
    - Sports
    - Culture
    - Nature
    - Events
- Item Features:
    - Item ID
    - Item Name
    - Ontological Category
- Ratings:
    - User ID
    - Item ID
    - Rating

Concerning the type of feature, they can be divided essentially into three groups: numerical, categorical ordinal and categorical nominal.

Concerning numerical and categorical ordinal features, we have the following:

- Numerical
    - Age – numerical (can be transformed into age bins)
- Ordinal:
    - Age bins = ['18-30','31-40', '41-50', '51-60', '60+']

- Academic Degree = ['None', 'High School', 'Some College', 'College Degree']
- Budget = ['Low', 'Mid', 'High']
- Accommodation = ['Single', 'Double', 'Suite', 'Villa']

As for categorical nominal features, the following were modelled:

- Gender = ['Male', 'Female']
- Job = ['Blue Collar', 'White Collar']
- Country/Region = ['South Europe', 'North Europe', 'East Europe', 'North America', 'South America', 'Asia', 'Africa', 'Middle East']
- Group Composition = ['1 Adult', '2 Adults', '2 Adults + Child', 'Group of Friends']

These features were selected to make up the synthetic dataset. The next steps start with the definition of how these features correlate with each other, and how to generate data sets with different types of features, that is, numerical, ordinal and nominal.

## 3.2 Step-by-step methodology application

### 3.2.1 Step 1)

In step 1), continuous correlated latent processes that correspond to the ordinal features are sampled. To this end, a covariance matrix, which contains the pairwise covariance values between features, is defined. This covariance matrix is the first input defined for the data generation and is presented in the following table. In this case, the correlation matrix is the same as the covariance matrix because all variance values (main diagonal) are equal to one.

*Table 1 Covariance matrix of the ordinal features*

|        | Age | AcDeg | Budget | Accom |
|--------|-----|-------|--------|-------|
| Age    | 1.0 | 0.4   | 0.5    | 0.5   |
| AcDeg  | 0.4 | 1.0   | 0.6    | 0.4   |
| Budget | 0.5 | 0.6   | 1.0    | 0.9   |
| Accom  | 0.5 | 0.4   | 0.9    | 1.0   |

The simulation of the Gaussian copula involves the following steps:

1. For an arbitrary covariance matrix, Σ, let P be its corresponding correlation matrix.
2. Compute the Cholesky decomposition, A, of P so that $P = A^T A$.
3. Generate $Z \sim MN_d(0, I_d)$.
4. Set $X = A^T Z$.
5. Return $U = (\Phi(X_1), \dots, \Phi(X_d))$.

In the present case, the correlation matrix, P, is equal to the covariance matrix which is the one presented in the previous table. For step 3 of the simulation of the Gaussian copula, each of the features is sampled with a uniform distribution between 0 and 1, after which the inverse of the cumulative distribution function (CDF) of the Gaussian distribution is used to map the sampled values, that correspond to cumulative probabilities, to standard normal distributed random variables, Z. In step 4 of the simulation, the sampled Gaussian random variables (Z) are multiplied by the Cholesky decomposition of the correlation matrix after which the correlated Gaussian random variables, X, are obtained.

In the following figures, an example of step 3 of the simulation of the Gaussian copula is shown.

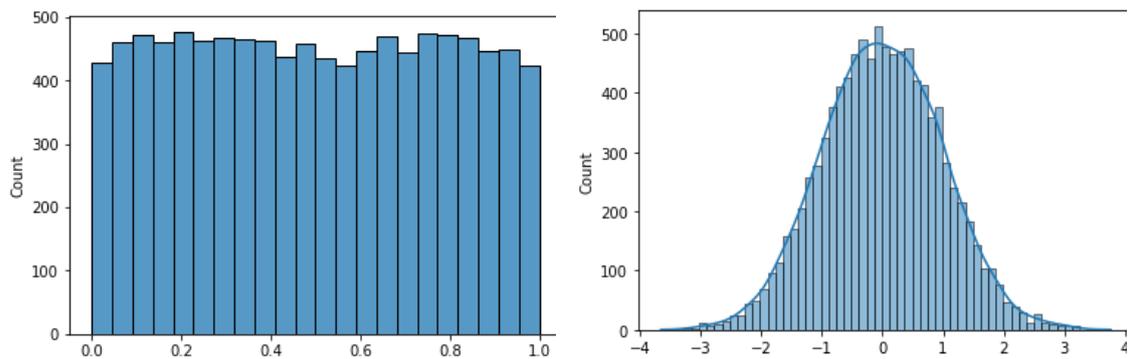

Figure 2 Example of step3 of the simulation of Gaussian copulas.

In the following figure, a sample of X is shown, showing two of the features, Budget and Academic Degree (AcDeg). In this figure the correlation is noticeable from the diagonal trend of the joint probability distribution.

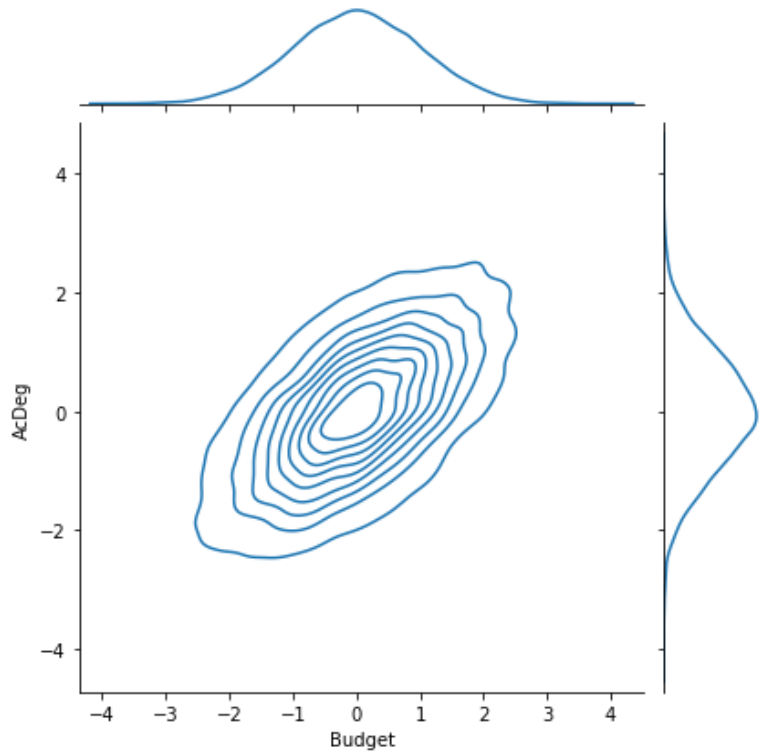

*Figure 3 Correlated features, AcDeg and Budget, resulting from the application of the Gaussian copulas with the previously defined correlation matrix.*

The Pearson correlation is calculated from the sampled random variables, X, confirming that the initial correlation matrix, P, corresponds (approximately) to the actual correlation matrix of the sampled random variables. The actual correlation matrix is presented in the table below, and by comparing with the correlation matrix, P, from Table 1, it is immediately apparent that the values are very close.

*Table 2 Actual Pearson correlation matrix calculated from the sampled random variables, X.*

|        | Age   | AcDeg | Budget | Accom |
|--------|-------|-------|--------|-------|
| Age    | 1.000 | 0.397 | 0.500  | 0.500 |
| AcDeg  | 0.397 | 1.000 | 0.601  | 0.398 |
| Budget | 0.500 | 0.601 | 1.000  | 0.899 |
| Accom  | 0.501 | 0.398 | 0.899  | 1.000 |

### 3.2.2 *Step 2)*

In step 2) of the data generation methodology, the α values related to the ordinal categories are defined. The α values relate to the probability of a given ordinal category and they correspond to the maximum value of the standard normal distribution (sampled standard normal variables) for which the feature takes that given categorical value.

The chosen α values are presented in the following tables, as well as a visualization of the concept of α values using the age bin case, where the α values in the corresponding table are the values in the x-axis of the figure, and the areas in different colours correspond to the probability of the corresponding age bin.

Table 3 Chosen α values and probabilities for the age bin categories

| Age bins | 18-29 | 30-39 | 40-49 | 50-59 | 60+ |
|---|---|---|---|---|---|
| α | -1.1 | -0.5 | 0.4 | 0.9 | - |
| Probability (%) | 13.6 | 17.3 | 34.7 | 16.0 | 18.4 |

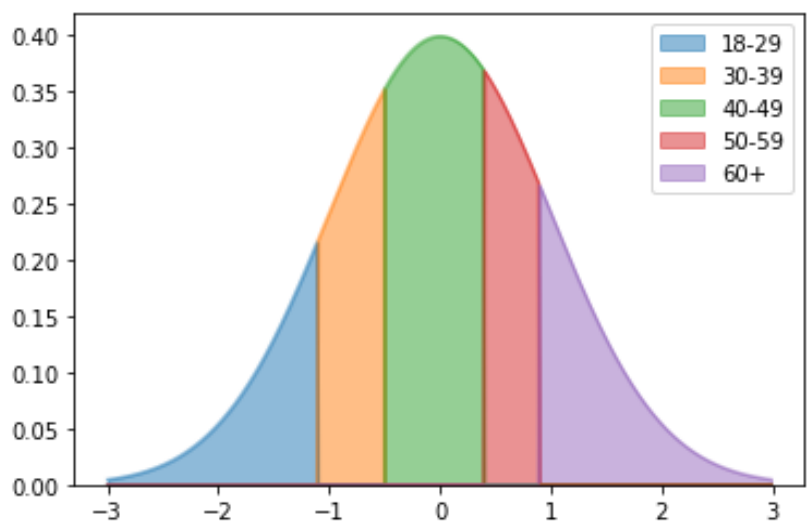

Figure 4 α values (from the table above) intersecting the standard normal probability density function of the continuous latent variable representing age. The areas of the graph correspond to the probability values given in the table above.

Table 4 Chosen α values and probabilities for the academic degree categories

| Academic Degree | None | High School | Some College | College Degree |
|---|---|---|---|---|
| α | -1.8 | -1.0 | 0.5 | - |
| Probability (%) | 3.6 | 12.3 | 53.3 | 30.8 |

Table 5 Chosen α values and probabilities for the budget categories

| Budget | Low | Mid | High |
|---|---|---|---|
| α | -0.5 | 1.2 | - |
| Probability (%) | 30.9 | 57.6 | 11.5 |

*Table 6 Chosen α values and probabilities for the accommodation categories*

| Accomodation | Single | Double | Suite | Villa |
|---|---|---|---|---|
| α | -1.0 | 1.0 | 1.7 | - |
| Probability (%) | 15.9 | 68.3 | 11.4 | 4.4 |

The following table shows a sample of the resulting dataset fraction where each category has been label encoded with an ordinal encoder from 1 to the cardinality of each feature.

*Table 7 Sample of the generated dataset fraction*

| UserID | Age | AcDeg | Budget | Accom |
|---|---|---|---|---|
| 0 | 4 | 2 | 1 | 2 |
| 1 | 5 | 4 | 2 | 3 |
| 2 | 3 | 3 | 2 | 2 |
| 3 | 4 | 4 | 2 | 2 |
| 4 | 3 | 3 | 2 | 3 |
| ... | ... | ... | ... | ... |
| 99995 | 4 | 4 | 2 | 2 |
| 99996 | 3 | 4 | 3 | 2 |
| 99997 | 1 | 1 | 1 | 1 |
| 99998 | 1 | 3 | 1 | 2 |
| 99999 | 4 | 3 | 2 | 2 |

### 3.2.3 Step 3) and 4)

In step 3) of the data generation methodology, the $α_d$ values (not to be confused with the previous α values) of each category that compose each nominal feature are defined. These $α_d$ values define the Dirichlet distribution which is the prior of the Multinomial distribution used to sample the nominal features. In step 4), the Multinomial distribution is sampled with the pre-defined Dirichlet priors for each of the nominal features. With this, the nominal feature of each user is assigned the value of the multinomial sample experiment whose probabilities of each outcome is given by the Dirichlet prior distribution obtained from the initial $α_d$ values defined in step 3).

In the following the $α_d$ values that were used are presented and the concept is better explained so that the reader gains some intuition. The $α_d$ values represent an amount of observations belonging to each category. These $α_d$ values are used to derive likely Dirichlet distributions given the observed results. Then a Dirichlet distribution is sampled to be used as prior for the Multinomial distribution, from which a sample is taken to define the category of a user for the respective nominal feature. In the following tables the $α_d$ values are shown for each category of each feature. In most cases the features are independent, however, in the case of the job feature categories, these were defined as correlated with the academic degree ordinal feature. The way this is modelled is by defining different $α_d$ values for different observed categories of the academic degree. This can be seen in Table 9 where with an academic degree of "None" it is much more likely to have an observation of "Blue Collar" than "White Collar" given, respectively, by the 90-10 ratio of observations. Conversely, with an academic degree category of "College Degree" the ratio is inverted with 10-90 in favour of the "White Collar" category. This is the intuition behind the $α_d$ values.

Table 8 $α_d$ values for the gender categories

| Categories gender | Male | Female |
|---|---|---|
| $α_d$ | 10 | 10 |

Table 9 $α_d$ values for the job categories. In this case, the job categories were defined as being correlated with the academic degree feature, with different $α_d$ values depending on the category of the academic degree.

| Categories job/ Academic Degree | Blue Collar | White Collar |
|---|---|---|
| $α_d$ – None | 90 | 10 |
| $α_d$ – High School | 70 | 30 |
| $α_d$ – Some College | 40 | 60 |
| $α_d$ – College Degree | 10 | 90 |

Table 10 $α_d$ values for the region categories

| Categories region | South Europe | North Europe | East Europe | North America | South America | Asia | Africa | Middle East |
|---|---|---|---|---|---|---|---|---|
| $α_d$ | 50 | 50 | 10 | 10 | 10 | 20 | 20 | 10 |

Table 11 $α_d$ values for the group composition categories

| Categories group composition | 1 Adult | 2 Adults | 2 Adults + Child | Group of friends |
|---|---|---|---|---|
| $α_d$ | 20 | 50 | 50 | 10 |

In the following table, a sample of the resulting data set is shown, where to the already previously defined ordinal features the sampled nominal features are added.

Table 12 Sample of the dataset including ordinal and nominal features.

| UserID | Age | AcDeg | Budget | Accom | Gender | Job | Region | GroupComp |
|---|---|---|---|---|---|---|---|---|
| 0 | 4 | 2 | 1 | 2 | Female | blue collar | North Europe | 2Adlt |
| 1 | 5 | 4 | 2 | 3 | Male | white collar | North Europe | GrpFriends |
| 2 | 3 | 3 | 2 | 2 | Female | blue collar | North Europe | 2Adlt+Child |
| 3 | 4 | 4 | 2 | 2 | Female | white collar | North Europe | 2Adlt+Child |
| 4 | 3 | 3 | 2 | 3 | Female | white collar | South Europe | 2Adlt |
| ... | ... | ... | ... | ... | ... | ... | ... | ... |
| 99995 | 4 | 4 | 2 | 2 | Female | white collar | North Europe | 2Adlt+Child |
| 99996 | 3 | 4 | 3 | 2 | Male | white collar | Asia | 2Adlt+Child |
| 99997 | 1 | 1 | 1 | 1 | Female | blue collar | South Europe | 2Adlt |
| 99998 | 1 | 3 | 1 | 2 | Female | blue collar | South Europe | 2Adlt+Child |
| 99999 | 4 | 3 | 2 | 2 | Male | blue collar | North America | 2Adlt+Child |

### 3.2.4 Step 5)

In step 5), the Multinomial Logit model is defined and applied to obtain the preferences of each user. The expressions presented before concerning the Multinomial Logit model were employed in this step. However, first the β values that define the marginal utilities of each feature on each preference have to be defined. These marginal utilities define a β matrix where each entry represents the weight that a user characteristic/feature has on a given preference.

In the following, the β matrix is presented, and the pseudo-code for the application of the Multinomial Logit is given. The marginal utilities are on a scale of 1 to 5, where 1 means that the marginal utility increases the least as the feature value increases, in the case of ordinal features, and means that the marginal utility is low for a given category in the case of nominal features. Conversely, 5 means that the marginal utility increases the most as the feature value increases, in the case of ordinal features, and means that the marginal utility is high for a given category in the case of nominal features.

Table 13 β matrix of marginal utilities of each feature/category per each preference category.

| Feature/Preference | Beach | Relax | Shop | Nightlife | Theme park | Gastro | Sports | Culture | Nature | Events |
|---|---|---|---|---|---|---|---|---|---|---|
| Age | 2 | 5 | 3 | 1 | 2 | 4 | 1 | 4 | 3 | 3 |
| Academic Degree | 3 | 4 | 3 | 3 | 3 | 3 | 3 | 4 | 3 | 3 |
| Budget | 3 | 3 | 5 | 3 | 3 | 4 | 3 | 3 | 3 | 3 |
| Accommodation | 3 | 3 | 3 | 3 | 3 | 3 | 3 | 3 | 3 | 3 |
| x0_Female | 3 | 3 | 4 | 3 | 3 | 3 | 3 | 3 | 3 | 3 |
| x0_Male | 3 | 3 | 3 | 3 | 3 | 3 | 4 | 3 | 3 | 3 |
| x1_Blue Collar | 3 | 3 | 3 | 4 | 4 | 1 | 4 | 2 | 3 | 3 |
| x1_White Collar | 3 | 3 | 4 | 3 | 3 | 5 | 3 | 5 | 3 | 3 |
| x2_Africa | 3 | 3 | 5 | 4 | 3 | 3 | 3 | 3 | 3 | 4 |
| x2_Asia | 2 | 3 | 4 | 2 | 4 | 3 | 2 | 4 | 3 | 3 |
| x2_EastEurope | 3 | 3 | 4 | 4 | 3 | 3 | 4 | 3 | 3 | 3 |
| x2_MiddleEast | 1 | 3 | 5 | 2 | 3 | 4 | 3 | 3 | 2 | 3 |
| x2_NorthAmerica | 4 | 3 | 3 | 3 | 4 | 3 | 4 | 3 | 4 | 3 |
| x2_NorthEurope | 4 | 3 | 3 | 4 | 3 | 4 | 5 | 4 | 5 | 3 |
| x2_SouthAmerica | 5 | 3 | 3 | 4 | 4 | 3 | 4 | 3 | 3 | 3 |
| x2_SouthEurope | 5 | 3 | 2 | 4 | 3 | 5 | 4 | 4 | 3 | 3 |
| x3_1Adult | 3 | 3 | 3 | 4 | 1 | 4 | 4 | 3 | 3 | 3 |
| x3_2Adults | 4 | 4 | 4 | 4 | 2 | 3 | 3 | 3 | 3 | 3 |
| x3_2Adults+Child | 4 | 5 | 3 | 1 | 4 | 3 | 3 | 3 | 3 | 3 |
| x3_GroupFriends | 3 | 3 | 4 | 5 | 4 | 4 | 3 | 3 | 4 | 4 |

In the following, a pseudocode of the Multinomial Logit model is provided:

**Function** *SimulateDatasetMultinomialLogit (**x, beta, norm_pref_id=0**):*

    *# Normalize β matrix according to chosen preference (default first preference "Beach")*

    ***beta** = normalizeBeta (**beta, norm_feat_id=0**)*

    *# Matrix multiplication between user-feature matrix and β matrix to obtain partial utilities, V*

    ***V** = **x** @ **beta***

    *# Sample random values from the Gumbel distr. for the idiosyncratic portion of the utilities, E*

    ***uni** = randomSample (0,1) (**dim(V)**)*

    ***E** = GumbelInverseCDF (**uni**)*

    *# Sum partial utilities to obtain utility matrix*

    ***U** = **V** + **E***

    *# Return β matrix and utility matrix*

    *return (**beta,U**)*

### 3.2.5 Step 6)

In step 6), the user preference vectors, or user preference matrix, are obtained from the results of the Multinomial Logit model defined in the previous step. The output of the Multinomial Logit is the utility matrix which comprises the normalized utilities that each user assigns to each of the preference categories.

To attain the preference vectors, the previously presented expression is applied to the utility vectors:

$$Pr(d|u_{ij}) = \frac{\exp(u_{id})}{\sum_{j=1}^{J} \exp(u_{ij})}$$

The resulting user-preference (U$_{Pref}$) matrix is attained and a sample of said matrix is presented in the following table.

Table 14 Sample of the user-preference (U$_{Pref}$) matrix.

| UserID | Beach | Relax | Shop | Nightlife | Theme park | Gastro | Sports | Culture | Nature | Events |
|---|---|---|---|---|---|---|---|---|---|---|
| 0 | 0.408 | 0.026 | 0.020 | 0.041 | 0.002 | 0.002 | 0.004 | 0.009 | 0.487 | 0.002 |
| 1 | 0.002 | 0.077 | 0.017 | 0.015 | 0.009 | 0.457 | 0.041 | 0.271 | 0.107 | 0.002 |
| 2 | 0.554 | 0.156 | 0.039 | 0.041 | 0.027 | 0.010 | 0.021 | 0.015 | 0.135 | 0.003 |
| 3 | 0.005 | 0.038 | 0.012 | 0.000 | 0.003 | 0.252 | 0.003 | 0.674 | 0.009 | 0.002 |
| 4 | 0.002 | 0.229 | 0.003 | 0.001 | 0.000 | 0.137 | 0.001 | 0.623 | 0.000 | 0.002 |
| ... | ... | ... | ... | ... | ... | ... | ... | ... | ... | ... |
| 99995 | 0.003 | 0.106 | 0.202 | 0.000 | 0.020 | 0.115 | 0.005 | 0.202 | 0.337 | 0.010 |
| 99996 | 0.001 | 0.127 | 0.064 | 0.000 | 0.002 | 0.034 | 0.001 | 0.750 | 0.016 | 0.005 |
| 99997 | 0.050 | 0.285 | 0.030 | 0.337 | 0.110 | 0.006 | 0.091 | 0.019 | 0.015 | 0.057 |
| 99998 | 0.031 | 0.712 | 0.007 | 0.083 | 0.103 | 0.004 | 0.021 | 0.027 | 0.006 | 0.007 |
| 99999 | 0.005 | 0.880 | 0.064 | 0.000 | 0.035 | 0.000 | 0.009 | 0.003 | 0.002 | 0.003 |

### 3.2.6 *Step 7) and 8)*

In step 7) a list of items is defined, with corresponding categories and an item category (icat) matrix is built.

Table 15 Item list and corresponding item category (link to ontology).

| itemID | Item Name | Category |
|---|---|---|
| 0 | Restaurant Fake | [Gastro] |
| 1 | Fiction Nightclub | [Nightlife] |
| 2 | Random Shopping Mall | [Shop, Relax] |
| 3 | Bogus Waterpark | [Theme park] |
| *4* | Unknown Nature Route | [Nature, Relax] |
| *5* | Some Sport Event | [Sports, Events] |
| *6* | Never Happened Festival | [Events, Culture] |
| *7* | False Tavern | [Gastro, Culture] |
| *8* | Make-believe Pub | [Gastro, Nightlife] |
| *9* | Another Sport Event | [Sports, Events] |
| *10* | Surprise Concert | [Events, Culture] |
| *11* | Museum of Fake History | [Culture] |
| *12* | Fake BTT Route | [Sports, Nature] |
| *13* | Random Surfing Lessons | [Sports, Nature] |
| *14* | Fake Brands Boutique | [Shop] |
| *15* | Best Imaginary Restaurant | [Gastro] |
| *16* | Bogus Spa | [Relax] |
| *17* | Random Cultural Tour | [Culture] |
| *18* | Non existing Zipline | [Sports] |
| *19* | Fake Klub | [Nightlife] |
| *20* | Random Golf Lessons | [Sports] |
| *21* | Secret Beach | [Beach] |
| *22* | Fake Beach | [Beach] |

The next part of this step is the vectorization of the above table. Then, in step 8), the $U_{Pref}$ matrix is multiplied with the vectorized $I_{Cat}$ matrix, thus obtaining the UserItem matrix. In the next table, the vectorized $I_{Cat}$ matrix is shown.

*Table 16 Vectorized item-category ($I_{Cat}$) matrix.*

| itemID | Beach | Relax | Shop | Nightlf | Themeprk | Gastro | Sports | Culture | Nature | Events |
|---|---|---|---|---|---|---|---|---|---|---|
| *0* | 0 | 0 | 0 | 0 | 0 | 1 | 0 | 0 | 0 | 0 |
| *1* | 0 | 0 | 0 | 1 | 0 | 0 | 0 | 0 | 0 | 0 |
| *2* | 0 | 1 | 1 | 0 | 0 | 0 | 0 | 0 | 0 | 0 |
| *3* | 0 | 0 | 0 | 0 | 1 | 0 | 0 | 0 | 0 | 0 |
| *4* | 0 | 1 | 0 | 0 | 0 | 0 | 0 | 0 | 1 | 0 |
| *5* | 0 | 0 | 0 | 0 | 0 | 0 | 1 | 0 | 0 | 1 |
| *6* | 0 | 0 | 0 | 0 | 0 | 0 | 0 | 1 | 0 | 1 |
| *7* | 0 | 0 | 0 | 0 | 0 | 1 | 0 | 1 | 0 | 0 |
| *8* | 0 | 0 | 0 | 1 | 0 | 1 | 0 | 0 | 0 | 0 |
| *9* | 0 | 0 | 0 | 0 | 0 | 0 | 1 | 0 | 0 | 1 |
| *10* | 0 | 0 | 0 | 0 | 0 | 0 | 0 | 1 | 0 | 1 |
| *11* | 0 | 0 | 0 | 0 | 0 | 0 | 0 | 1 | 0 | 0 |
| *12* | 0 | 0 | 0 | 0 | 0 | 0 | 1 | 0 | 1 | 0 |
| *13* | 0 | 0 | 0 | 0 | 0 | 0 | 1 | 0 | 1 | 0 |
| *14* | 0 | 0 | 1 | 0 | 0 | 0 | 0 | 0 | 0 | 0 |
| *15* | 0 | 0 | 0 | 0 | 0 | 1 | 0 | 0 | 0 | 0 |

| | 0 | 1 | 2 | 3 | 4 | 5 | 6 | 7 | 8 | 9 |
|---|---|---|---|---|---|---|---|---|---|---|
| 16 | 0 | 1 | 0 | 0 | 0 | 0 | 0 | 0 | 0 | 0 |
| 17 | 0 | 0 | 0 | 0 | 0 | 0 | 0 | 1 | 0 | 0 |
| 18 | 0 | 0 | 0 | 0 | 0 | 0 | 1 | 0 | 0 | 0 |
| 19 | 0 | 0 | 0 | 1 | 0 | 0 | 0 | 0 | 0 | 0 |
| 20 | 0 | 0 | 0 | 0 | 0 | 0 | 1 | 0 | 0 | 0 |
| 21 | 1 | 0 | 0 | 0 | 0 | 0 | 0 | 0 | 0 | 0 |
| 22 | 1 | 0 | 0 | 0 | 0 | 0 | 0 | 0 | 0 | 0 |

As mentioned, the multiplication of the $U_{Pref}$ matrix with the $I_{Cat}$ matrix gives origin to the UserItem matrix, which translates the users' preferences concerning the item categories to the items *per se*.

$$[UserItem] = [U_{Pref}] \times [I_{Cat}]$$

The values are mostly between 0 and 1 and can be interpreted as the probability of a given user liking an item. A sample of the UserItem matrix is presented in the following table.

*Table 17 Sample of the UserItem matrix*

| | 0 | 1 | 2 | 3 | 4 | 5 | ... | 17 | 18 | 19 | 20 | 21 | 22 |
|---|---|---|---|---|---|---|---|---|---|---|---|---|---|
| 0 | 0.002 | 0.041 | 0.046 | 0.002 | 0.513 | 0.006 | ... | 0.009 | 0.004 | 0.041 | 0.004 | 0.408 | 0.408 |
| 1 | 0.457 | 0.015 | 0.094 | 0.009 | 0.184 | 0.043 | ... | 0.271 | 0.041 | 0.015 | 0.041 | 0.002 | 0.002 |
| 2 | 0.010 | 0.041 | 0.195 | 0.027 | 0.291 | 0.024 | ... | 0.015 | 0.021 | 0.041 | 0.021 | 0.554 | 0.554 |
| 3 | 0.252 | 0.000 | 0.050 | 0.003 | 0.048 | 0.006 | ... | 0.674 | 0.003 | 0.000 | 0.003 | 0.005 | 0.005 |
| 4 | 0.137 | 0.001 | 0.233 | 0.000 | 0.230 | 0.003 | ... | 0.623 | 0.001 | 0.001 | 0.001 | 0.002 | 0.002 |
| ... | ... | ... | ... | ... | ... | ... | ... | ... | ... | ... | ... | ... | ... |
| 99995 | 0.115 | 0.000 | 0.308 | 0.020 | 0.442 | 0.015 | ... | 0.202 | 0.005 | 0.000 | 0.005 | 0.003 | 0.003 |
| 99996 | 0.034 | 0.000 | 0.191 | 0.002 | 0.143 | 0.006 | ... | 0.750 | 0.001 | 0.000 | 0.001 | 0.001 | 0.001 |
| 99997 | 0.006 | 0.337 | 0.315 | 0.110 | 0.300 | 0.148 | ... | 0.019 | 0.091 | 0.337 | 0.091 | 0.050 | 0.050 |
| 99998 | 0.004 | 0.083 | 0.718 | 0.103 | 0.717 | 0.028 | ... | 0.027 | 0.021 | 0.083 | 0.021 | 0.031 | 0.031 |
| 99999 | 0.000 | 0.000 | 0.943 | 0.035 | 0.882 | 0.012 | ... | 0.003 | 0.009 | 0.000 | 0.009 | 0.005 | 0.005 |

### 3.2.7 *Step 9) and 10)*

In step 9) noise is added to the user-item preference matrix to introduce some further randomness and uncertainty to the preferences, and consequently, by propagation, to the ratings. Some uncertainty had already been inserted via the use of the Gumbel distribution and the insertion of extreme values in the utility calculation for the multinomial logit model. The noise was inserted through the sum of a random sparse matrix generated with a predefined level of sparsity/density, this is different to the sparsity measure, which measures the sparsity of the ratings matrix, in this case the sparsity refers to the density of the random sparse noise matrix that is added to the user-item preference matrix. In this instance, the chosen density of the random sparse noise matrix was 1%.

Finally, in step 10) the ratings are generated using a Fuzzy Inference System. To run the rating generator module, besides defining a set of fuzzy rules, one has to define user behavioural traits for each user, and also perceived implicit item quality. In addition to this, the sparsity measure also must be defined beforehand. As previously mentioned, this sparsity measure corresponds to the density of the user-item ratings matrix, since this is a sparse matrix where most of the entries are empty. One of the advantages of generating one's synthetic data is that different user-item ratings matrix density values can be tested in order to test the recommendation algorithms at different stages of recommender system maturity.

To start this last step, one starts with the user behavioural traits and the perceived implicit item quality. These are essential for the Fuzzy Inference System because they are inputs of the fuzzy rule base. Intuitively, when a user rates an item, the rating will not only reflect a given user's preference, but rather it will be a mix of that factor and a few others, such as the perceived quality of the item and the way a user rates the items, i.e., the user's behaviour. Regarding the perceived quality of the item, this is intuitive since, for example, even if someone really enjoys Mexican restaurants that doesn't mean that person will always rate every Mexican restaurant highly. He or she will likely rate good Mexican restaurants well and bad ones not so well. Concerning user behaviour, i.e., how a user rates items, this has to do with a different issue. It is not unexpected that two persons that equally liked the same item, let's say a restaurant, give unequal ratings. One can observe this by noticing that user average ratings are not the same across users, some users rate on average higher than others. In addition, not all users have the same rating dispersion, this means that it is plausible to find some users that might, on a 1-5 scale, rate the majority of the items with values from 2-4, only giving out 5s and 1s in extreme cases, and a different user that rates mainly with the extreme values 1 and 5. These different aspects should be encoded in the data so that the dataset reflects these real-world behaviours. To that end, three latent features are defined, one pertaining to item quality, and two pertaining to user average rating and rate dispersion. These three features, 'quality', 'bias', 'spread', are encoded as numerical features from 1 to 5 in the case of 'quality' and 'bias' and 1 to 4 for 'spread' and are randomly generated inside the given value intervals for each item (quality) and user (bias, spread). These numerical latent features are then fuzzified along with the user-item preference values calculated in step 8).

*Table 18 Sample of DataFrame containing user bias and user spread*

|       | userId | bias | spread |
|-------|--------|------|--------|
| 0     | 0      | 3.59 | 1.71   |
| 1     | 1      | 4.71 | 1.68   |
| 2     | 2      | 2.94 | 3.29   |
| 3     | 3      | 3.54 | 1.01   |
| 4     | 4      | 1.09 | 1.58   |
| ...   | ...    | ...  | ...    |
| 99995 | 99995  | 1.83 | 1.88   |
| 99996 | 99996  | 3.22 | 1.12   |
| 99997 | 99997  | 1.23 | 1.29   |
| 99998 | 99998  | 3.48 | 2.12   |
| 99999 | 99999  | 2.49 | 3.17   |

The fuzzification is performed, mapping the numerical values to categories of item quality, user bias and user spread. In the following figures, the fuzzy sets with the labels to which the numerical values are fuzzified are shown. To implement the Fuzzy Inference System, a Python toolbox for fuzzy logic was used called scikit-fuzzy (skfuzzy) [30].

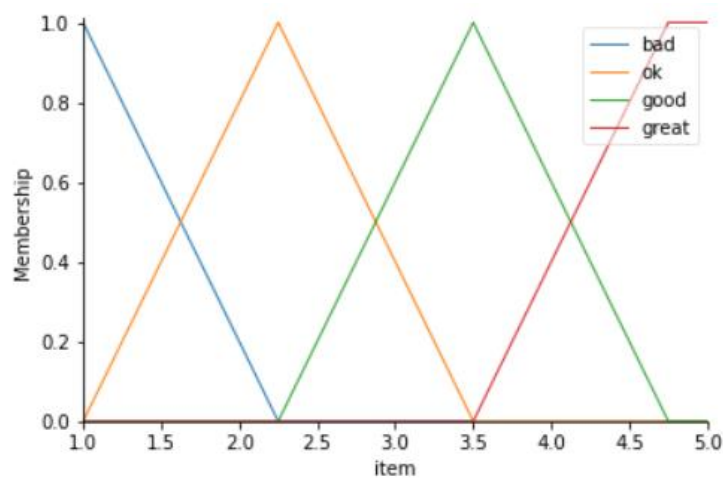

*Figure 5 Fuzzy set for item quality.*

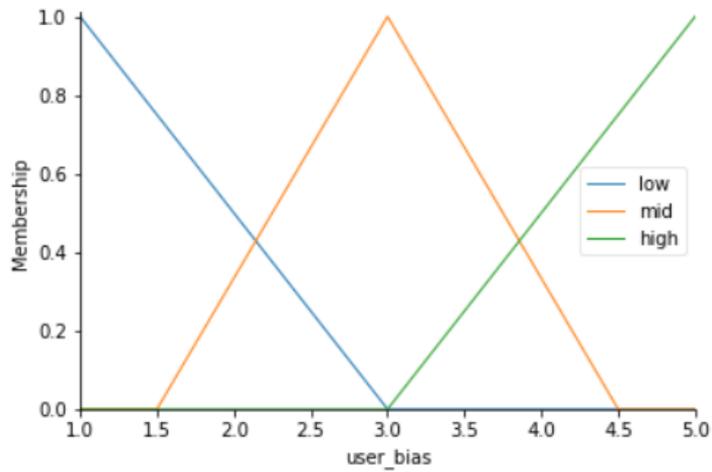

*Figure 6 Fuzzy set for user bias.*

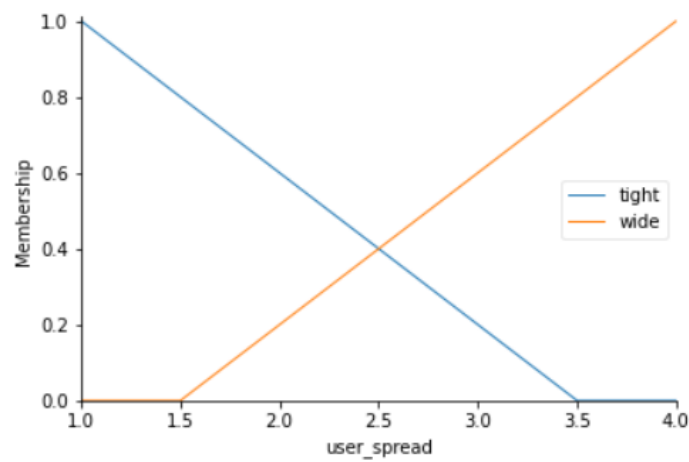

*Figure 7 Fuzzy set for user spread.*

The Mamdani approach is applied to map the inputs to the output of the Fuzzy Inference System. For this to happen to more fuzzy sets have to be introduced, one concerning the user preferences, and another related to the output for the Mamdani approach. The output is defined by a set of fuzzy rules that map the interaction of the fuzzy inputs to the fuzzy output. In the following, the fuzzy sets are presented as well as a sample of the fuzzy rules.

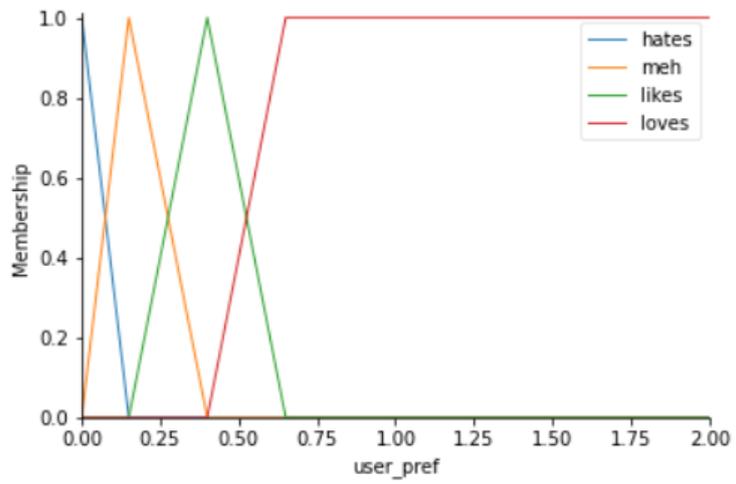

*Figure 8 Fuzzy set for user preference.*

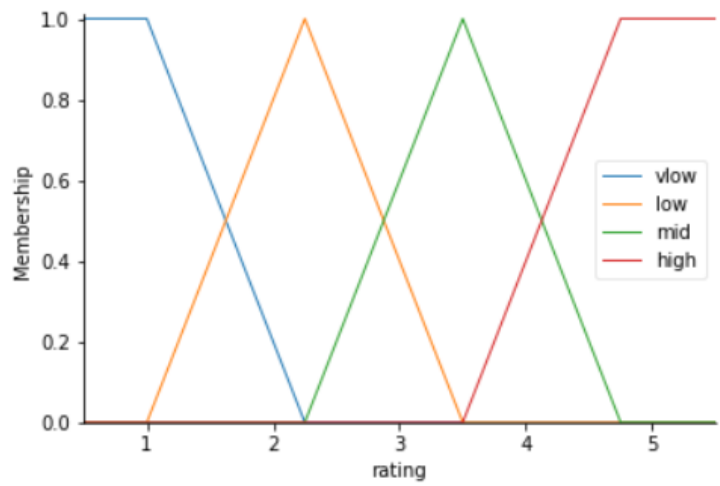

*Figure 9 Fuzzy set for the output of the Fuzzy Inference System, which corresponds to user rating.*

```python
R1 = ctrl.Rule(user_pref['hates'] & user_spread['tight'] & user_bias['low'] & item['bad'], rating['vlow'])
R2 = ctrl.Rule(user_pref['hates'] & user_spread['tight'] & user_bias['low'] & item['ok'], rating['vlow'])
R3 = ctrl.Rule(user_pref['hates'] & user_spread['tight'] & user_bias['low'] & item['good'], rating['vlow'])
R4 = ctrl.Rule(user_pref['hates'] & user_spread['tight'] & user_bias['low'] & item['great'], rating['low'])

R5 = ctrl.Rule(user_pref['hates'] & user_spread['tight'] & user_bias['mid'] & item['bad'], rating['vlow'])
R6 = ctrl.Rule(user_pref['hates'] & user_spread['tight'] & user_bias['mid'] & item['ok'], rating['vlow'])
R7 = ctrl.Rule(user_pref['hates'] & user_spread['tight'] & user_bias['mid'] & item['good'], rating['low'])
R8 = ctrl.Rule(user_pref['hates'] & user_spread['tight'] & user_bias['mid'] & item['great'], rating['low'])

R9 = ctrl.Rule(user_pref['hates'] & user_spread['tight'] & user_bias['high'] & item['bad'], rating['vlow'])
R10 = ctrl.Rule(user_pref['hates'] & user_spread['tight'] & user_bias['high'] & item['ok'], rating['low'])
R11 = ctrl.Rule(user_pref['hates'] & user_spread['tight'] & user_bias['high'] & item['good'], rating['low'])
R12 = ctrl.Rule(user_pref['hates'] & user_spread['tight'] & user_bias['high'] & item['great'], rating['mid'])

R13 = ctrl.Rule(user_pref['hates'] & user_spread['wide'] & user_bias['low'] & item['bad'], rating['vlow'])
R14 = ctrl.Rule(user_pref['hates'] & user_spread['wide'] & user_bias['low'] & item['ok'], rating['vlow'])
R15 = ctrl.Rule(user_pref['hates'] & user_spread['wide'] & user_bias['low'] & item['good'], rating['vlow'])
R16 = ctrl.Rule(user_pref['hates'] & user_spread['wide'] & user_bias['low'] & item['great'], rating['vlow'])

R17 = ctrl.Rule(user_pref['hates'] & user_spread['wide'] & user_bias['mid'] & item['bad'], rating['vlow'])
R18 = ctrl.Rule(user_pref['hates'] & user_spread['wide'] & user_bias['mid'] & item['ok'], rating['vlow'])
R19 = ctrl.Rule(user_pref['hates'] & user_spread['wide'] & user_bias['mid'] & item['good'], rating['vlow'])
R20 = ctrl.Rule(user_pref['hates'] & user_spread['wide'] & user_bias['mid'] & item['great'], rating['mid'])

R21 = ctrl.Rule(user_pref['hates'] & user_spread['wide'] & user_bias['high'] & item['bad'], rating['vlow'])
R22 = ctrl.Rule(user_pref['hates'] & user_spread['wide'] & user_bias['high'] & item['ok'], rating['vlow'])
R23 = ctrl.Rule(user_pref['hates'] & user_spread['wide'] & user_bias['high'] & item['good'], rating['mid'])
R24 = ctrl.Rule(user_pref['hates'] & user_spread['wide'] & user_bias['high'] & item['great'], rating['mid'])
```

*Figure 10 Sample of the user rules when user preference equals the label 'hates'.*

The way the rules are read is as follows, using as an example rule R1: If the user 'hates' the item category, and the user's rating spread is 'tight' and the user tends to rate 'low' (user_bias), and the item is perceived as 'bad', then the rating value given by the user should be very low ('vlow'). A total of 96 rules to cover all possible combinations of parameters were defined. With the rules defined the ratings matrix can be calculated by running the Fuzzy Inference System, which takes the fuzzy rules and fuzzy sets as input, as well as the user-item preference matrix, information about item quality, user behavior and finally density of the random sparse noise matrix (defined as 1% for this case study) and the density of the user-item ratings matrix (defined as 15% for this case study). In the following table, a sample of the attained user-item ratings matrix is presented, where zero values represent unrated items.

*Table 19 Sample of the calculated user-item ratings matrix.*

| userId | 0 | 1 | 2 | 3 | 4 | 5 | ... | 17 | 18 | 19 | 20 | 21 | 22 |
|---|---|---|---|---|---|---|---|---|---|---|---|---|---|
| 0 | 0.00 | 1.50 | 0.00 | 0.00 | 0.00 | 0.00 | ... | 2.83 | 0.00 | 0.00 | 0.00 | 1.63 | 0.00 |
| 1 | 0.00 | 0.00 | 0.00 | 0.00 | 0.00 | 4.16 | ... | 0.00 | 0.00 | 0.00 | 0.00 | 0.00 | 0.00 |
| 2 | 0.00 | 0.00 | 2.71 | 0.00 | 0.00 | 0.00 | ... | 0.00 | 0.00 | 0.00 | 0.00 | 0.00 | 0.00 |
| 3 | 0.00 | 0.00 | 0.00 | 0.00 | 0.00 | 0.00 | ... | 0.00 | 0.00 | 0.00 | 0.00 | 0.00 | 2.66 |

| | | | | | | | | | | | | |
|---|---|---|---|---|---|---|---|---|---|---|---|---|
| **4** | 0.00 | 0.00 | 1.37 | 0.00 | 0.00 | 4.71 | ... | 0.00 | 0.00 | 0.00 | 3.80 | 1.18 | 0.00 |
| **...** | ... | ... | ... | ... | ... | ... | ... | ... | ... | ... | ... | ... | ... |
| **99995** | 1.87 | 0.00 | 0.00 | 0.00 | 0.00 | 3.24 | ... | 0.00 | 0.00 | 0.00 | 0.00 | 0.00 | 0.00 |
| **99996** | 0.00 | 0.00 | 0.00 | 0.00 | 0.00 | 0.00 | ... | 2.44 | 1.73 | 0.00 | 0.00 | 0.00 | 0.00 |
| **99997** | 0.00 | 0.00 | 0.00 | 0.00 | 0.00 | 0.00 | ... | 0.00 | 0.00 | 0.00 | 0.00 | 0.00 | 0.00 |
| **99998** | 0.00 | 0.00 | 0.00 | 0.00 | 0.00 | 0.00 | ... | 0.00 | 0.00 | 0.00 | 0.00 | 0.00 | 0.00 |
| **99999** | 0.00 | 0.00 | 0.00 | 0.00 | 0.00 | 2.89 | ... | 0.00 | 0.00 | 2.89 | 0.00 | 0.00 | 2.89 |

Some statistical properties of the obtained ratings matrix are presented next. First, it is interesting to observe the actual density of the sparse matrix. The output of that is 0.1502, which corresponds to the 15% value that was defined as input. Then some simple descriptive statistics, with mean, minimum and maximum values, and a histogram.

*Table 20 Mean, minimum and maximum values in the user-item ratings matrix.*

| | |
|---|---|
| *Mean* | 2.49 |
| *Minimum* | 1.13 |
| *Maximum* | 4.76 |

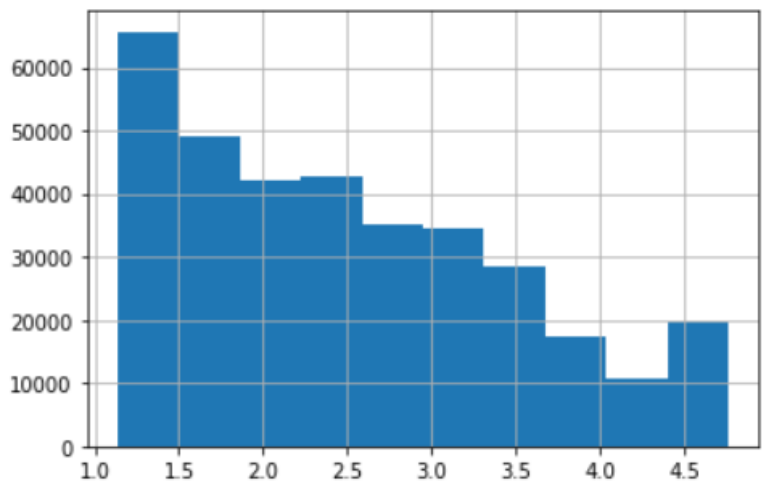

*Figure 11 Histogram of the sparse user-item ratings matrix. The x-axis corresponds to rating value and the y-axis to observed frequency.*

# 4  Conclusions

The developed methodology allows to obtain datasets for machine learning tasks, and particularly for recommender systems, which are composed of a mixture of datatypes: continuous, ordinal, and nominal, via two different strategies, one for continuous and ordinal features, and another for nominal features. The methodology also generates preferences related to the user and item features via a multinomial logit model. In the last step of the methodology a Fuzzy Inference System is employed to generate user ratings of the items, based on the user's preferences, behavioural traits, item categories and implied quality. The author believes that the methodology is novel in terms of the strategy for the generation of nominal features and the subsequent employment of a multinomial logit model for the definition of preferences and finally with the use of a Fuzzy Inference System to generate the ratings. In addition, there aren't many works focusing on the generation of datasets with multiple datatypes in which a strategy for dealing with the "correlation" between nominal and ordinal features. There are also not many works on the generation of synthetic data for recommender systems. Therefore, the author believes that this work brings a novel contribution in the field when compared with the available literature. The use of this work and its applicability is also worth mentioning. In fact, the methodology tries to give an answer to a serious generalized issue concerning lack of good quality datasets for data science projects, and in particular ones concerning the development and training of recommender systems. As was mentioned along the manuscript, nowadays, regardless of the ubiquitous nature of data, it is very common either not to have enough data or very bad quality data, or legally not being able to use the data that one has, or the data not having all the necessary features for our project, or even not having any data at all. In any of these cases, this methodology can be useful. Another important aspect of this work is that with synthetic data one is able to modify certain characteristics, be it distributions or introducing some sort of data drift to perform some kind of sensitivity analysis. In addition, with a synthetic dataset, the user knows which latent features were used to generate the user preferences. This is important because with real datasets, this is not readily known, and may not be observable if the features collected don't have any weight on user preferences. The implication is that with synthetic datasets one is actually able to verify whether or not certain algorithms work correctly on the data. For example, applying Field-Aware Factorization Machines to observe the latent factors associated to the different features in the dataset, which then can be compared with the beta values from the Multinomial Logit model.

## 4.1  Future Works

As future works it would make sense to eventually introduce other techniques of synthetic dataset generation such as Generative Adversarial Networks (GANs). Also, the inclusion of a module that allows to insert data drift into some features. This would make sense in a module where a time feature also exists. This module would allow for the generation of timeseries data. Another addition would be the possibility of using different fuzzy sets and adding different rating profiles, such as users that rate more than others. Additionally, it would also make sense to allow different distributions of the rated items, from random selection (as is implemented here) where the items rated are randomly selected to giving a larger likelihood of selection to items that belong to categories/preferences that the user likes. The development of a platform where all these possible scenarios are included is in the horizon, and where a user of the platform would identify their needs, be it for recommender systems or other data science projects.

## 5    Acknowledgments

The present paper was developed in the context of the PMP project – Partnership Management Platform, code LISBOA-01-0247-FEDER-045411, co-financed by LISBOA 2020 and Portugal 2020 through the European Regional Development Fund.